\documentclass[proceedings]{JHEP} 

\newcommand{\be}{\begin{equation}}
\newcommand{\ee}{\end{equation}}
\newcommand{\bea}{\begin{eqnarray}}
\newcommand{\eea}{\end{eqnarray}}
\newcommand{\ba}{\begin{array}}
\newcommand{\ea}{\end{array}}

\def\npb#1#2#3{Nucl.\ Phys.\ {\bf B#1}, #2 (#3)}
\def\prl#1#2#3{Phys.\ Rev.\ Lett.\ {\bf #1}, #2 (#3)}
\def\prd#1#2#3{Phys.\ Rev.\ {\bf D#1}, #2 (#3)}
\def\plb#1#2#3{Phys.\ Lett.\ {\bf B#1}, #2 (#3)}
\def\zpc#1#2#3{Z.\ Phys. {\bf C#1}, #2 (#3)}
\def\moda#1#2#3{Mod.\ Phys.\ Lett.\ {\bf A#1}, #2 (#3)}
\def\jhep#1#2#3{JHEP \ {\bf #1}, #2 (#3)}

\def\lsim{\mathrel{\hbox{\rlap{\hbox{\lower4pt\hbox{$\sim$}}}\hbox{$<$}}}}
\def\gsim{\mathrel{\hbox{\rlap{\hbox{\lower4pt\hbox{$\sim$}}}\hbox{$>$}}}}
 
\usepackage{epsfig}

\conference{Corfu Summer Institute on Elementary Particle Physics, 1998}

\title{Effective Supergravity from Heterotic M--Theory and its 
                                  Phenomenological Implications}

\author{Carlos Mu\~noz\thanks{I thank
   K.Choi, H.B. Kim and D.G. Cerde\~no for their collaboration in
this project.}\\
        Departamento de F\'{\i}sica Te\'orica C-XI and Instituto de
F\'{\i}sica Te\'orica C-XVI\\
Universidad Aut\'onoma de Madrid, 28049 Cantoblanco, Madrid, Spain

        E-mail: \email{carlos.munnoz@uam.es}}

\abstract{
In this talk I summarize several recent results concerning the
four--dimensional effective
supergravity obtained using a Calabi--Yau compactification 
of the $E_8\times E_8$ heterotic string from M--theory. 
A simple 
macroscopic study is provided 
expanding the theory
in powers of two dimensionless variables.
Higher order terms in the K\"ahler potential are identified  and
matched with the heterotic string corrections. 
In the context of this M--theory expansion, I discuss several phenomenological
issues:
universality of soft scalar
masses,
relations between  
the different scales of the theory
(eleven--dimensional Planck mass, compactification scale and orbifold
scale) in order to obtain 
unification at   
$3\times 10^{16}$ GeV or lower values,
soft supersymmetry--breaking
terms, and finally charge and colour breaking minima.
The above analyses are also carried out 
in the presence of 
(non--perturbative) 
five--branes.

\bigskip

~~~~~~~~~~~~~~~~~~~~~~~~~~~~~~~~~~~~~~~~~~~~~~~~~~~~~~~~~~~~~~~~~~~~~~~~~~~~~~~~~~~~~~~~~~~~~~~~~~~~~~~~~~~~~~~~~~~~~~~~~~~~~~~~~~~~~~~~~~~~~~~~~~~~~~~~~~~~~~~~~~~~~~~~~~~~~~~~~~~~~~~~~~~~~~~~~~~~~~~~~~~~~~~~~~~~~~~~~~~~~~~~~~~~~~~~~~~~~

{F FTUAM 99/19} \\
{IFT-UAM/CSIC-99-24} \\
{June 1999}

}

\begin{document}

\section{Introduction and summary}

One of the most exciting proposals of the last years in 
string theory, 
consists of the
possibility that the five distinct superstring theories in ten
dimensions plus
supergravity in eleven dimensions  
be different vacua in the moduli space of a single underlying 
eleven--dimensional 
theory, the so--called M--theory \cite{Schwarz}.
In this respect,
Ho\v{r}ava and Witten proposed that
the strong--coupling limit of  $E_8\times E_8$ heterotic
string theory can be obtained from M--theory. They used the low--energy
limit of M--theory, 
eleven-dimensional
supergravity, on  a manifold with boundary (a $S^1/Z_2$ orbifold), 
with the 
$E_8$ gauge multiplets at each of the 
two ten--dimensional boundaries (the orbifold fixed planes) 
\cite{Horava-Witten}.

In the present paper I will summarize several recent results concerning
the four--dimensional implications of this so called heterotic M--theory. 
In particular, I will concentrate on the analysis of the effective 
supergravity obtained by compactifying 
heterotic M--theory on a six--dimensional Calabi--Yau manifold 
and its phenomenological consequences.

The effective action of this  limit 
has been  systematically analyzed in an expansion in powers of
$\kappa^{2/3}$, where $\kappa^2$ denotes the eleven--dimensional
gravitational coupling \cite{Witten}.
As was noticed in \cite{Banks-Dine}, this leads to 
an expansion parameter which  scales as $\kappa^{2/3}\rho/V^{2/3}$,
where $\pi\rho$ denotes the length of the eleventh segment
and $V$ is the Calabi--Yau volume.
At the leading order in this expansion,
the K\"ahler potential, superpotential and gauge kinetic functions
have been computed in
\cite{Banks-Dine,Li-Lopez-Nanopoulos1,Dudas-Grojean,%
Nilles-Olechowski-Yamaguchi}.
It is rather easy to determine 
the order $\epsilon_1$ correction to the leading
order gauge kinetic functions
\cite{Banks-Dine,Choi,Nilles-Olechowski-Yamaguchi,Nilles-Stieberger},
while it is much more nontrivial to compute
the order $\epsilon_1$ correction to
the leading order K\"ahler potential,
which was recently done by  Lukas, Ovrut and Waldram 
\cite{Lukas-Ovrut-Waldram}.
It was argued in \cite{Mio} that
the holomorphy and Peccei--Quinn symmetries guarantee
that there is {\it no} further correction
to the gauge kinetic functions and the superpotential 
at any finite order
in the M--theory expansion, similarly to the case of the
perturbative heterotic string \cite{Nillesloop}.

On the other hand, as is well known, 
the four--dimensional effective action of the weakly
coupled heterotic string theory can be expanded in powers of the
two dimensionless variables: the string coupling
$\epsilon_s=e^{2\phi}/(2\pi)^5$ 
and the worldsheet sigma--model coupling
$\epsilon_{\sigma}=4\pi\alpha^{\prime}/V^{1/3}$. 
It was suggested in \cite{Mio} that 
the effective action of M-theory can be similarly analyzed 
by expanding it in powers of the two dimensionless variables:
$\epsilon_1=\kappa^{2/3}\pi\rho/V^{2/3}$ and $\epsilon_2=
\kappa^{2/3}/\pi\rho V^{1/3}$.
The latter
is the straightforward
generalization of the string world--sheet 
coupling $\sim \alpha^{\prime}/V^{1/3}$ to the membrane
world--volume coupling $\sim \kappa^{2/3}/\rho V^{1/3}$
since $\kappa^{2/3}$ may be
identified as the inverse of the membrane tension.
Note that in the M--theory limit, heterotic string corresponds to 
a membrane stretched along the eleventh dimension.
In this framework the K\"ahler potential is expected to 
receive corrections which are higher order in $\epsilon_1$ or 
$\epsilon_2$.
An explicit computation of these  higher order corrections
will be highly nontrivial  
since first of all  the eleven--dimensional action is known
only up to the terms of order $\kappa^{2/3}$ relative to the zeroth
order action (except for the order $\kappa^{4/3}$ four-gaugino term)
and secondly the higher order computation of the compactification solution
and its  Kaluza-Klein reduction
are  much more complicated.

In section~2, we will provide a simple macroscopic analysis
of the four--dimensional effective supergravity action by expanding it 
in powers of $\epsilon_1$ and $\epsilon_2$.
Possible higher order corrections in the K\"ahler potential
are identified and matched with the heterotic string corrections,
and their size is estimated for the physically 
interesting values of moduli \cite{Mio}.
The validity of this 
procedure has been explicitly checked in \cite{Wyllard} in the
case of M--theory compactified on
$S^1/Z_2\times K3\times T^2$.

On the other hand, we will also discuss in detail 
how these effective supergravity
models can be strongly constrained
by imposing the phenomenological requirement of universal soft 
scalar masses, in order to avoid 
dangerous flavour changing neutral current phenomena.
As pointed out in \cite{five-branes2}, there is a simple solution 
to avoid this problem:
to work  with Calabi--Yau spaces with one K\"ahler modulus $T$ only.
Of course, the existence of such spaces, as e.g. the quintic hypersurface in
$CP^4$, and their universality properties 
was also known in the context of the weakly--coupled heterotic
string,
however the novel fact in heterotic M--theory, 
is that model building  
is relatively easy.
For example, in the presence of non--standard embedding and
five--branes (non--perturbative objects located at points throughout
the orbifold interval \cite{Witten}) to obtain three--generation models with
realistic gauge groups, as for example $SU(5)$, is not specially
difficult \cite{Donagi}.

Other
phenomenological implications of heterotic M--theory,
turn out to be also advantageous with respect
to the ones of the perturbative heterotic--string theory.
First of all, 
the resulting four--dimensional effective theory can reconcile 
the observed Planck scale $M_{Planck}= 1.2 \times 10^{19}$ GeV 
with the phenomenologically favored GUT scale 
$M_{GUT}\approx 3\times 10^{16}$
GeV in a natural manner, providing an attractive framework
for the unification of couplings \cite{Witten,Banks-Dine}.
This is to be compared to the weakly--coupled heterotic string where
$M_{string}\approx 8.5\times 10^{17}$ GeV.
Another phenomenological virtue of the M--theory limit
is that there can be a QCD axion whose high energy axion potential is
suppressed enough so that the  strong CP problem can be solved 
by the axion mechanism \cite{Banks-Dine,Choi}.
About the issue of supersymmetry 
breaking, the possibility of generating it by the
gaugino condensation on the hidden boundary has been studied
\cite{Horava,Nilles-Olechowski-Yamaguchi,Lalak-Thomas,Lukas-Ovrut-Waldram2,Quiros,Choi-Kim-Kim}
and also some interesting features of the
resulting soft supersymmetry--breaking terms were discussed. 
In particular, gaugino masses turn
out to be of the same order as squark 
masses \cite{Nilles-Olechowski-Yamaguchi}. This is welcome since
gaugino
masses much smaller than squark masses, as in the     
weakly--coupled heterotic string case, 
may give rise to a hierarchy
problem \cite{Beatriz}. 
For example, the experimental lower bound on gluino masses 
of $150$ GeV would imply
scalar masses larger than $1$ TeV. 
Besides, the phenomenologically favored vacuum expectation values of
the moduli can be obtained with several gaugino condensates with 
the appropriate hidden matter \cite{Choi-Kim-Kim}, similarly to the
case of the weakly--coupled heterotic string \cite{condensates}.
However, it is fair to say that unlike the latter non--perturbative
membrane instantons are also necessary in M--theory to obtain
the desired minimum. 

Several of the above mentioned phenomenological issues will be analyzed in
the next sections. 
In section~3 we will concentrate in the case of standard and
non--standard
embedding 
vacua \cite{Stieberger,Benakli,Lalak,five-branes}, 
whereas in section~4 vacua in the presence of
five--branes \cite{five-branes,five-branes2} are studied.
The latter are characterized basically by new moduli $Z_n$
associated with the five--brane positions in the orbifold dimension.
In both cases we will perform a 
detailed study of the
different scales of the theory,  
as well as a 
systematic
analysis
of the soft supersymmetry--breaking terms.

Concerning the former,
the relations between the eleven--dimensional Planck mass, the
Calabi--Yau compactification
scale and the orbifold scale, taking into account higher order
corrections
to the leading order formulae, will be analyzed \cite{Mio2}.
Identifying 
the compactification scale with the GUT scale, 
to obtain $M_{GUT} \approx 3\times 10^{16}$ GeV is simpler in 
non--standard embedding models than in standard ones. 
In the presence of five--branes, $M_{GUT}$ can be obtained more
easily.
On the other hand, 
going away from perturbative vacua, it was
recently realized that the 
string scale may be anywhere between the weak scale and the Planck
scale \cite{Lykken} 
and the size of the extra dimensions 
may be as large as a millimetre \cite{Dimopoulos}.
Whether or not all these scenarios\footnote{             
To trust them would imply to assume that 
Nature is trying to mislead us with an apparent gauge coupling 
unification at $M_{GUT}$. 
In this sense, a reasonable doubt about those possibilities is
healthy.} 
are possible in the context of 
heterotic M--theory
has been analyzed recently in \cite{Benakli2} with interesting results:
to lower the unification scale 
(and therefore the eleven--dimensional Planck scale which is around two times
bigger)
to intermediate values $\approx 10^{11}$ GeV or $1$ TeV values or 
to obtain the radius of the orbifold as large as a millimetre 
is in principle
possible in some special limits.  
However, it has been pointed out
in \cite{Mio2} that the necessity of a fine--tuning  
or the existence of a hierarchy problem 
renders these possibilities
unnatural.
Although new possibilities arise in the presence of five--branes 
in order to lower the scales, 
again at the cost of introducing a huge hierarchy problem.

We will also analyze
the soft supersymmetry breaking terms under the general
assumption that supersymmetry is spontaneously broken by
the auxiliary components of the bulk moduli superfields 
(dilaton $S$ and modulus $T$) \cite{Mio,Lukas-Ovrut-Waldram2,Lii,Mio2}. 
It is examined in particular how the soft terms
vary when one moves from the weakly--coupled heterotic string
limit to the strongly--coupled limit. The presence of
new parameters in the formulae gives rise to different pattern of soft
terms. This is also the case of models with five--branes where
at least a new goldstino angle, associated to a modulus
$Z_1$, must be included in the computation
\cite{five-branes2,Kubo,Mio2}.
Unlike the weakly--coupled case, scalar masses larger
than gaugino masses can be obtained \cite{Mio2}.
Low--energy 
($\approx M_W$) sparticle 
spectra \cite{Lopez,Mio,Bailin,Kawamura,Mio2}
are also discussed.

Finally, the existence of charge and colour breaking minima is
analyzed. 
As is well known, the presence of scalar fields with colour and 
electric charge in supersymmetric theories induces the possible
existence of dangerous charge and colour breaking minima, which would
make the standard vacuum unstable \cite{CCB}.
They impose very strong constraints on supergravity models
from heterotic M--theory \cite{Savoy,Alejandro}.
In particular, standard--embedding models turn out to be excluded
on these grounds, similarly
to the perturbative heterotic-string situation \cite{Alejandro}.
Possible solutions to this problem are discussed.

\section{Four--dimensional effective supergravity}

Here we will analyze the four--dimensional effective supergravity
obtained by compactifying heterotic M--theory on a six--dimensional
Calabi--Yau manifold.

\subsection{Expansions}

Let us first discuss possible perturbative expansions
of the four--dimensional effective supergravity.
As in the case of the weakly coupled heterotic string theory,
the effective supergravity of compactified M--theory
contains two model--indepen\-dent moduli
superfields $S$ and $T$ whose scalar components can be identified as 
\bea
{\rm Re}(S) &=& \frac{1}{2\pi (4\pi)^{2/3}} M_{11}^{6} V\ ,
\nonumber \\
{\rm Re}(T) &=& \frac{6^{1/3}}{2(4\pi)^{4/3}} M_{11}^{3} V^{1/3} \pi\rho\ ,
\label{dilaton-modulus}
\eea
with $M_{11}$ denoting the
eleven--dimensional
Planck mass, $\kappa^{2}=M_{11}^{-9}$.
The above normalizations of $S$ and $T$ have been chosen to keep the
conventional form of the gauge kinetic functions in the effective supergravity.
(See (\ref{gaugef}) for our form of the gauge kinetic functions.
Our $S$ and $T$ correspond to $\frac{1}{4\pi}S$ and $\frac{1}{8\pi}T$ 
of \cite{Choi} respectively.)

The moduli $S$ and $T$ can be used to define
various kind of  expansions which may be applied for
the low--energy effective action. For instance, in the weakly
coupled heterotic string limit,
we have
\bea
{\rm Re}(S) &=& e^{-2\phi}\frac{V}{(2\alpha^{\prime})^3}\ ,
\nonumber \\
{\rm Re}(T) &=& \frac{6^{1/3}}{32\pi^3}\frac{V^{1/3}}{2\alpha^{\prime}}\ ,
\eea
where $\phi$ and $\sqrt{2\alpha^{\prime}}$ denote the heterotic
string dilaton and length scale respectively.
One may then expand the effective action of the heterotic string
theory  in powers of the string
loop expansion parameter $\epsilon_s$ and the world--sheet
sigma model expansion parameter $\epsilon_{\sigma}$:   
\bea
&&\epsilon_s = \frac{e^{2\phi}}{(2\pi)^5}
\approx 0.3 \frac{[4\pi^2 {\rm Re}(T)]^3}{{\rm Re}(S)}\ , 
\nonumber \\
&&\epsilon_{\sigma} = \frac{4\pi\alpha^{\prime}}{V^{1/3}}
\approx 0.5 \frac{1}{4\pi^2{\rm Re}(T)}\ .
\label{quepasa}
\eea
Here we are interested in the possible expansion in the M--theory limit of 
the strong heterotic--string coupling $\epsilon_s\gg 1$ 
for which  $\pi\rho\gsim M_{11}^{-1}$ and $V\gsim M_{11}^{-6}$ and  so
the physics can be described by eleven--dimensional supergravity. 
Since we have two independent length scales, $\rho$ and $V^{1/6}$, 
there can be two dimensionless expansion
parameters in the M--theory limit also. 
As discussed in the introduction there are two natural candidates,
$\epsilon_1$ and $\epsilon_2$, to
be the expansion parameters of 
the four--dimensional effective supergravity action of the Ho\v{r}ava-Witten
M--theory. Using $\kappa^{2}=M_{11}^{-9}$ these can be written as
\bea
&&\epsilon_1 = \frac{\pi\rho}{M_{11}^3 V^{2/3}}\approx 
 \frac{{\rm Re}(T)}{{\rm Re}(S)}\ , 
\nonumber \\
&&\epsilon_2=\frac{1}{M_{11}^3 \pi\rho V^{1/3}}\approx 
\frac{1}{4\pi^2 {\rm Re}(T)}\ ,
\eea
where (\ref{dilaton-modulus}) has been  used  to arrive at this expression of
$\epsilon_1$ and $\epsilon_2$. 
Note that $\epsilon_1\epsilon_2\approx 1/[4\pi^2 {\rm Re}(S)]\\ \approx
\alpha_{GUT}/\pi$ which is essentially the four dimensional field
theory expansion parameter.
Thus  if one goes to the limit in which one
expansion works better while keeping the realistic value
of $\alpha_{GUT}$,
the other expansion becomes worse.
Here we will simply assume that both $\epsilon_1$ and $\epsilon_2$
are small enough so that  the double expansion in $\epsilon_1$
and $\epsilon_2$  provides
a good perturbative scheme for the effective action
of M--theory. 
As we will see later, it turns out that this expansion
works well even when $\epsilon_1$ becomes of order one,
which is in fact  necessary to have $M_{GUT}\approx 3\times
10^{16}$ GeV.

To be explicit, let us consider a simple compactification on  
a Calabi-Yau manifold with the Hodge-Betti number
$h_{1,1}=1$. 
In this model, the low--energy degrees of freedom include first
the gravity multiplet and $S$ and $T$ which are 
the massless modes of
the eleven--dimensional bulk fields. 
We also have gauge and charged matter superfields associated
to the observable and hidden sector gauge groups, 
$G_O\times G_H \subset E_8\times E_8$, 
where $G_O$($G_H$) is located at the boundary 
$x^{11}=0$($x^{11}=\pi\rho$) with $x^{11}$ denoting the orbifold
coordinate. From now on, we will use as our
notation the subscript $O$($H$) for quantities and functions of the
observable(hidden) sector. 

It is then easy to compute the K\"ahler potential $K$, the
observable and hidden sector gauge kinetic functions
$f_{O}$ and $f_{H}$, and the superpotential $W$
at the leading order in the M--theory expansion. 
Obviously the leading contribution to the
moduli K\"ahler metric is from 
the  eleven--dimensional bulk field  action 
which is of order $\kappa^{-2}$, 
while the
charged matter K\"ahler metric,  the gauge kinetic functions, and the
charged matter superpotential receive the leading contributions
from the ten--dimensional boundary action which is
of order $\kappa^{-4/3}$.
One finds
\cite{Banks-Dine,Li-Lopez-Nanopoulos1,Dudas-Grojean,Nilles-Olechowski-Yamaguchi}
\bea
&& K=
-\ln (S+\bar{S})-3\ln (T+\bar{T})+\frac{3}{T+\bar{T}}C^p {\bar C}^p
\nonumber \\
&& f_{O}=f_{H}=S\ , 
\nonumber \\
&& W=d_{pqr}C^pC^qC^r\ ,
\label{tree}
\eea
where $d_{pqr}$ are constant coefficients and $C^p$ are the matter
fields, i.e. the effective supergravity computed at the leading
order in the M--theory expansion is the same as the effective
supergravity of the 
weakly--coupled heterotic
string computed at the leading order in the string loop and sigma
model perturbation theory. 

The holomorphy and the Peccei-Quinn symmetries  
imply that there is no correction to  the superpotential
at any finite order in the $S$ and $T$--dependent
expansion parameters $\epsilon_1$ and $\epsilon_2$.
However the gauge kinetic functions can receive a correction at order
$\epsilon_1$ in a way  consistent with the holomorphy and the Peccei-Quinn
symmetries.
This correction 
can be determined by a direct M--theory computation \cite{Witten}
or by matching the string loop threshold correction
to the gauge kinetic function 
\cite{Banks-Dine,Choi,Nilles-Stieberger,Nilles-Olechowski-Yamaguchi}.
The result is
\be
f_{O}=S+\beta_O T\ , \quad f_{H}=S+\beta_H T\ , 
\label{gaugef}
\ee
where the model--dependent integer coefficients
$\beta_{O,H}={1\over 8\pi^2}\int\omega\wedge[{\rm tr}
(F_{O,H}\wedge F_{O,H})-\frac{1}{2}{\rm tr}(R\wedge R)]$, 
for the K\"ahler form $\omega$ normalized as
the generator of the integer (1,1) 
cohomology\footnote{
Usually $\beta$ is considered to be an arbitrary real number.
For $T$ normalized as (\ref{dilaton-modulus}),
it is required to be an integer \cite{Choi}.}, and 
they 
fulfil the following
condition:
\bea
\beta_O + \beta_H &=& 0\ ,
\label{constraint}
\eea
with $\beta_O$ always positive in the case of the standard
embedding 
of the spin connection into one of the $E_8$ gauge groups. Positive and
negative
values are possible for non--standard embedding 
cases \cite{Benakli,Lalak,five-branes}.

Let us now consider the possible higher order corrections to
the K\"ahler potential.
With the Peccei-Quinn symmetries, the K\"ahler potential can be written as 
$K=\hat{K}(S+\bar{S},T+\bar{T})+Z(S+\bar{S},T+\bar{T})C^p {\bar C}^p$
with
$\hat{K}=\hat{K}_0+\delta \hat{K}$, $Z=Z_0+\delta Z$.
Here $\hat{K}_0=-\ln (S+\bar{S})-3\ln (T+\bar{T})$ and $Z_0=3/(T+\bar{T})$ 
denote the leading order 
results in (\ref{tree}), while $\delta\hat{K}$ and $\delta Z$
are the higher order corrections.
Before going to the M--theory expansion of 
$\delta\hat{K}$ and $\delta Z$, it is useful to note that
the bulk physics become blind to the existence of boundaries
in the limit $\rho\rightarrow \infty$.
However some of the boundary physics, e.g. the boundary Calabi-Yau
volume, can be affected by the integral of the bulk variables over
the eleventh dimension and then they can include a piece linear in 
$\rho$ \cite{Witten}.
This implies that $\delta\hat{K}/\hat{K}_0$, being the
correction to the pure bulk dynamics, contains only a non-negative
power of $1/\rho$ in the M--theory expansion,
while $\delta Z/ Z_0$ which concerns the couplings
between the bulk and boundary fields can include a piece  linear in $\rho$.
Since $\epsilon_1^n\epsilon_2^m\sim \rho^{n-m}$, one needs $m\geq n$
for the expansion of $\delta \hat{K}/\hat{K}_0$ and $m\geq n-1$
for the expansion of $\delta Z/Z_0$.
Taking account of these, the M--theory expansion
of the K\"ahler potential  is given by \cite{Mio}
\bea
\delta \hat{K}&=&\sum_{(n+m\geq 1, m\geq n)}   A_{nm}
\epsilon_1^n\epsilon_2^m 
=
\sum_{m\geq 1}\frac{A_{0m}}{[4\pi^2 {\rm Re}(T)]^m}
\nonumber\\ 
& + & \frac{A_{11}}{4\pi^2 {\rm Re}(S)}\left[ 1+{\cal O}(
\frac{1}{4\pi^2{\rm Re}(S)},\frac{1}{4\pi^2{\rm Re}(T)})\right]
\nonumber \\
\delta Z & = &\frac{3}{(T+\bar{T})}\sum_{(n+m\geq 1, m\geq n-1)}  B_{nm}
\epsilon_1^n\epsilon_2^m \nonumber \\
& = & \frac{3}{(T+\bar{T})}
\sum_{m\geq 1} \frac{B_{0m}}{[4\pi^2 {\rm Re}(T)]^m}
+\frac{3B_{10}}{2{\rm Re}(S)}
\nonumber\\ 
& \times &
\left[1+ {\cal O}(\frac{1}{4\pi^2
{\rm Re}(S)}, \frac{1}{4\pi^2 {\rm Re}(T)})\right]
\label{expansion}
\eea
where the $n=0$ terms are separated from the other
terms with $n\geq 1$.

The above expansion would work well 
in the M--theory limit: 
$[4\pi^2 {\rm Re}(T)]^3\gg {\rm Re}(S) \gg {\rm Re}(T)\\ \gg \frac{1}{4\pi^2}$,
while the heterotic string loop and sigma model
expansions work well in the heterotic string limit: 
${\rm Re}(S)\gg [4\pi^2 {\rm Re}(T)]^3, \quad {\rm Re}(T) \gg 
\frac{1}{4\pi^2}$.
By varying ${\rm Re}(S)$ while keeping ${\rm Re}(T)$
fixed, one can smoothly move 
from the M--theory limit $\epsilon_s\gg 1$ 
to the heterotic string limit $\epsilon_s\ll 1$  (or vice versa)
while keeping $\epsilon_1\approx {\rm Re}(T)/{\rm Re}(S)$ 
and $\epsilon_2\approx 1/[4\pi^2 {\rm Re}(T)]$  small enough.
Obviously then the M--theory K\"ahler potential expanded in
$\epsilon_1$ and $\epsilon_2$ remains to be valid over this
procedure, and thus is a  valid expression of the K\"ahler potential
even in the  heterotic
string limit.
This means that, like the case of the gauge kinetic functions,
one can determine the expansion coefficients 
in (\ref{expansion})
by matching  
the heterotic string K\"ahler potential which 
can be computed 
in the string loop and sigma model perturbation theory.
Since $\epsilon_1^n\epsilon_2^m\sim \epsilon_s^n
\epsilon_{\sigma}^{m+2n}$,   $(n,m)$-th order in the M--theory expansion
corresponds to $(n,m+2n)$-th order in the string loop and
sigma-model perturbation theory. 
Thus all the terms in the M--theory expansion
have their counterparts in the heterotic string expansion.
It appears that  the converse is not true in general,
for instance the term $\epsilon_s^p\epsilon_{\sigma}^q$ with $q<2p$
in the heterotic string expansion 
does  not have its counterpart in the M--theory expansion.
However all  string one--loop corrections
which have been computed so far
lead to corrections which scale
(relative to the leading terms) as  
$\epsilon_s\epsilon_{\sigma}^2$
or $\epsilon_s\epsilon_{\sigma}^3$, 
and thus have M--theory counterparts.
This leads us to suspect that all the terms that actually
appear in the heterotic
string expansion have $q\geq 2p$ and thus
have their counterparts in the M--theory  
expansion. Then there will be  a complete matching, 
up to (nonperturbative) corrections which
can not be taken into account by the M--theory expansion,
of the K\"ahler
potential between the M--theory limit and the heterotic string
limit,  like  the case of the gauge kinetic function
and superpotential. 
Collecting available informations  on the coefficients
in (\ref{expansion}),  either from the heterotic string analysis
or from the direct M--theory 
analysis (see \cite{Mio} and references therein)
one obtains the following higher order corrections 
to the leading
order K\"ahler potential in (\ref{tree}): 
%
\bea
\delta \hat{K} & = & \frac{A_{03}}{[4\pi^2 {\rm Re}(T)]^3}
\left[ 1+{\cal O}(
\frac{1}{4\pi^2{\rm Re}(T)})\right] 
\nonumber\\
& + &\frac{A_{11}}{4\pi^2 {\rm Re}(S)}
\left[ 1+{\cal O}(
\frac{1}{4\pi^2{\rm Re}(S)},\frac{1}{4\pi^2{\rm Re}(T)})\right]
\nonumber \\
\delta Z & = & 
\frac{3}{(T+\bar{T})}\frac{B_{03}}{[4\pi^2 {\rm Re}(T)]^3}
\left[ 1+{\cal O}(
\frac{1}{4\pi^2{\rm Re}(T)})\right] \nonumber \\
& + & \frac{\beta_O}{2{\rm Re}(S)}
\left[ 1+{\cal O}(
\frac{1}{4\pi^2{\rm Re}(S)},\frac{1}{4\pi^2{\rm Re}(T)})\right]
\nonumber\\
\label{correction}
\eea 
where $A_{03}, A_{11}$ and $B_{03}$ are of order one.

As a phenomenological application of the M--theory expansion
discussed so far, we are going to
analyze in subsection~3.2 the soft supersymmetry--breaking terms 
under the assumption that supersymmetry is spontaneously broken by
the auxiliary components $F^S$ and $F^T$ of the moduli superfields
$S$ and $T$.  
We will see in subsection~3.1 how moduli values of order one are necessary
in order to obtain $M_{GUT}\approx 3\times 10^{16}$ GeV.
Clearly, if ${\rm Re}(T)$ is of order one,
we are in the M--theory domain with  $\epsilon_s\gg 1$. (See (\ref{quepasa})).
One may  
worry  that the M--theory expansion (\ref{expansion}) would  not work
in this case
since 
$\epsilon_1={\rm Re}(T)/{\rm Re}(S)$ is of order one also.
However as we have noticed, any correction which is $n$-th order
in $\epsilon_1$ accompanies at least $(n-1)$-powers of
$\epsilon_2$ and thus is suppressed by $(\epsilon_1\epsilon_2)^{n-1}
\approx (\alpha_{GUT}/\pi)^{n-1}$ compared to
the order $\epsilon_1$ correction.
This allows the M--theory expansion (\ref{expansion}) to be valid even when
$\epsilon_1$ becomes of order one.
Obviously if ${\rm Re}(T)$ is of order one, 
only the order $\epsilon_1$ correction to $Z$,
i.e. $\delta Z=\beta_O/2{\rm Re}(S)$, can be sizable.
The other corrections are  suppressed  by  either 
$\epsilon_1\epsilon_2\approx 1/4\pi^2{\rm Re}(S)$ or $\epsilon_2^3
\approx 1/[4\pi^2 {\rm Re}(T)]^3$ and thus smaller than
the leading order results at least by  ${\cal O}(\frac{\alpha_{GUT}}{\pi})$.
Thus we will include  only $\delta Z=\beta_O/2{\rm Re}(S)$ 
($\delta Z=\beta_H/2{\rm Re}(S)$ for hidden matter) 
in the later
analysis of soft terms, while ignoring
the other corrections to the K\"ahler potential

Summarizing the above discussion, 
our starting point of the phenomenological
analyses in next sections
is the effective supergravity model given by
\bea
K &=& -\ln (S+\bar{S}) -3\ln (T+\bar{T})
\nonumber\\ 
& + & \frac{3}{T+\bar{T}}
\left(1+\frac{1}{3}\epsilon_O\right) C_O^p \bar{C}_O^p 
\nonumber\\ 
& + & \frac{3}{T+\bar{T}}
\left(1+\frac{1}{3}\epsilon_H\right) C_H^p \bar{C}_H^p \ ,
\label{kahler}
\\
f_{O} &=& S+\beta_O T\ , \quad f_{H}=S+\beta_H T\ , 
\label{kinetic}
\\
W_O &=& d_{pqr}C_O^pC_O^qC_O^r\ ,
\label{superpotential}
\eea
with
\be
\epsilon_O=\beta_O  
 \frac{T+\bar T}{S+\bar S}\ ,
\quad \epsilon_H=\beta_H  
 \frac{T+\bar T}{S+\bar S}\ .
\label{epsilonO}
\ee
Notice that the parameter $\epsilon_O$ defined above is
$\epsilon_O\approx \beta_O \epsilon_1$.
Here the superpotential and gauge kinetic functions are exact up to 
nonperturbative corrections, while there can be small additional  perturbative
corrections to the K\"ahler potential
which are of order $1/4\pi^2 {\rm Re}(S)$ or $1/[4\pi^2 {\rm Re}(T)]^3$.

\subsection{Universality of soft terms}

To carry out an exhaustive analysis of the phenomenology associated to
heterotic M-theory
compactified on a Calabi-Yau manifold one also should  
consider in principle models with more than one single $T$--modulus.
However, models with several moduli $T_i$ has the potential problem
of non--universal soft scalar masses \cite{Jerusalen}. 
The soft scalar  masses are given in general 
by 
${m}^2_{{p}{\bar{q}}} = 
m_{3/2}^2 Z_{p\bar q}- 
{F}^{m}\left( \partial_m\partial_{\bar{n}}
{Z_{{p}{\bar{q}}}}-  {Z}^{r\bar{s}}\partial_m {Z}_{p\bar{s}}
\partial_{\bar{n}}{Z}_{r\bar{q}}  \right){\bar F}^{\bar n}$ \cite{
Brignole-Ibanez-Munoz},
where 
$F^m=F^S, F^{T_i}$ in our case,
and $Z_{p\bar{q}}$ and
$Z^{p\bar{q}}$ denote the K\"ahler metric and its inverse
of the matter fields $C^p$. For example, in the case of the standard
embedding the K\"ahler metric of the matter fields $C^i$ is given by
$Z_{i\bar j} = ({\partial}^2 {\hat K}^T/\partial T_i \partial {\bar
  T}_j) e^{-{\hat K}^T/3} \\ + \delta Z_{i\bar j}(S+\bar{S},T_k+\bar{T}_k)$,
where 
${\hat K}^T=-\ln k_{ijk} (T_i+{\bar T}_i)(T_j+{\bar T}_j)(T_k+{\bar T}_k)$
%
and $\delta Z_{i\bar j}$ corresponds to
the $S$-dependent correction in the M--theory expansion (or
the string-loop correction). 
After normalizing the fields to get canonical kinetic terms, although the
first piece in $m_{i\bar j}^2$ above 
will lead to universal diagonal soft masses,
the second piece will generically induce non--universal contributions,
as in the case of the weakly--coupled limit of the heterotic string 
compactified on a Calabi--Yau \cite{Kim-Munoz},
due to the presence of the off--diagonal K\"ahler metric 
$Z_{i\bar j}$ written above.
This clearly implies that the scalar mass eigenvalues will be in 
general non--degenerate.
If one ignores $\delta Z_{i\bar j}$, the matter K\"ahler metric
is $S$-in\-de\-pen\-dent and, as a consequence,
in the dilaton--dominated
\cite{Kaplunovsky-Louis,Brignole-Ibanez-Munoz2} scenario with
$F^{T_i}=0$ 
the normalized soft scalar masses are   
universal as $m_i=m_{3/2}$. 
However including the $S$-dependent $\delta Z_{i\bar j}$,
one generically loses the scalar mass universality 
even in the dilaton--dominated case \cite{Mio}.
In fact, this 
was noted in \cite{Nir} 
for the string--loop induced\footnote{It is worth noticing
that supergravity--loop corrections may also 
induce non-universality \cite{Sik}.}
$\delta Z_{i\bar j}$
which is small in the weakly coupled heterotic string limit.
The main point here
is that in the M--theory limit $\delta Z_{i\bar j}$
can be as large as the 
leading order K\"ahler metric,
and then 
there can be a large violation
of the scalar mass universality even in the dilaton--dominated 
scenario \cite{Mio}. An explicit computation of $\delta Z_{i\bar j}$
can be found in \cite{multiLukas}.
Clearly, multimoduli Calabi--Yau models have the potential problem
of non--universal soft scalar masses.
Of course this can be ameliorated taking into account the low--energy
running of the scalar masses \cite{Brignole-Ibanez-Munoz2}. 
In particular, in the squark case, for
gluino masses heavier than (or of the same order as) the squark masses
at the boundary scale, there are large flavour-inde\-pen\-dent gluino
loop contributions which are the dominant source of squark masses.
However, to avoid the problem of non--universality from the beginning would be
welcome. As pointed out in \cite{five-branes2} there is the solution
of working with Calabi--Yau spaces with 
one K\"ahler modulus $T$ ($h_{1,1}=1$).
Clearly, supersymmetry breaking in the $S$ and/or $T$
direction in this case
will give rise to universal soft 
terms\footnote{
Although (2,1) complex structure moduli, $U_p$, may contribute
to the matter K\"ahler metric
with some $U$--dependent metric $H_{pq}$ multiplying the third and
forth term in (\ref{kahler}), they will not spoil the universality of
soft terms as long
as they do not contribute to supersymmetry breaking, $F^{U_p}=0$. I
thank A. Lukas and D. Waldram  
for useful discussions about this point.}.  

Notice that this improvement with respect to 
the problem of non--universality is
not possible in other compactifications. For example, although in most 
orbifolds the structure of soft scalar masses is simpler due to the
existence of diagonal metrics
$Z_{p{\bar q}}=\delta_{pq}\prod_i (T_i+\bar 
T_i)^{n_{p}^i}$, 
still they show a lack of universality $m_{p}^2=m_{3/2}^2+\sum_i \frac{n_{p}^i}
{(T_i+\bar T_i)^2}|F^{T_i}|^2$ \cite{Scheich}, due
to the modular weight dependence $n_{p}^i$ \cite{Lust-Ibanez}.
Although the above formulae are valid for the weakly--coupled case,
the result about non--universality of soft terms
is not modified in the strongly--coupled case \cite{Mio}.

Summarizing the above discussions, due to the constraints that the
universality of soft terms impose on effective supergravity models,
our starting point for the phenomenological analyses in the next section
is the model given by
(\ref{kahler}), (\ref{kinetic}) and (\ref{superpotential}),
i.e.
we will assume that the 
standard model arises from heterotic M--theory 
compactified on a Calabi--Yau
manifold with only one modulus field $T$.

\section{Phenomenology of standard and non--standard embedding vacua}

Here we will summarize first results found in the literature about the
standard and non--standard embedding cases, and then
we will 
discuss in detail the issue of the scales in the theory as well as
the pattern 
of 
soft terms. 

Let us recall first that the form of the effective action is 
determined by (\ref{kahler}), (\ref{kinetic}) and (\ref{superpotential}).
This is also true for the non--standard embedding case although there
is no requirement that the spin connection be embedded in the gauge 
connection \cite{Benakli,Lalak,five-branes}. 
Taking into account that the real parts of the gauge kinetic functions 
in (\ref{kinetic}) multiplied by $4\pi$ are the inverse gauge coupling
constants $\alpha_O$ and $\alpha_H$, using (\ref{dilaton-modulus})
one can write \cite{Witten,Li-Lopez-Nanopoulos1,Benakli}
\bea
\alpha_{O,H}
&=& \frac{(4\pi)^{2/3}}{2 M_{11}^{6} V_{O,H}}\ ,
\label{alphaO'}
\eea
with 
%
%
$V_{O,H}=V(1+\epsilon_{O,H})$
%
the observable(hidden) sector volume.
%
%
%
%
%

On the other hand, using $V_O$ as defined above,
the M-theory expression of the four--dimensional Planck scale 
%
$M_{Planck}^2 = 16\pi^2\rho M_{11}^9 <V>$
%
where $<V>$ is the average volume of the Calabi--Yau space
%
$<V>= (V_O+V_H)/2$,
%
and (\ref{dilaton-modulus}) one finds
\bea
V_O^{-1/6} 
&=&
\left(\frac{V}{<V>}\right)^{1/2}
3.6\times 10^{16} 
\left(\frac{4}{S+\bar S}\right)^{1/2}
\nonumber\\ 
& \times &
\left(\frac{2}{T+ \bar T}\right)^{1/2} 
\left(\frac{1}{1+\epsilon_O}\right)^{1/6}
{\rm GeV}
\ ,
\label{gut}
\eea
which is a very useful formula 
as we will see below in order to 
discuss whether or not the GUT scale or smaller scales 
are obtained in a natural way. In this respect,
let us now obtain the connection between the different scales of the
theory:
the eleven--dimensional Planck mass, $M_{11}$, the
Calabi--Yau compactification scale, $V_O^{-1/6}$,
and the orbifold scale, $(\pi \rho)^{-1}$. 
It is straightforward to obtain from (\ref{alphaO'}) the following
relation:
\bea
\frac{M_{11}}{V_O^{-1/6}} &=& 
2
\ .
\label{relation}
\eea
Likewise, using the above expression for $M_{Planck}$
and (\ref{alphaO'}) 
we arrive at
\bea
\frac{V_O^{-1/6}}{(\pi \rho)^{-1}} 
&=&
\left(\frac{V}{<V>}\right) 
\left(\frac{2.7\times 10^{16}{\rm GeV}}{V_O^{-1/6}}\right)^2 
\nonumber\\ 
& \times &
7\ (1+\epsilon_O)
\ .
\label{relation2}
\eea 
Notice that in (\ref{relation}) and (\ref{relation2})
we have already assumed that 
the gauge group of the observable sector
$G_O$ is the one of the standard model or some unification gauge group
as $SU(5)$, $SO(10)$ or $E_6$, i.e. we are using
$\left(2\pi\alpha_O\right)^{-1}=4$ in
order to reproduce the LEP data about 
$\alpha_{GUT}$ ($\alpha_O$ in our notation).

Let us recall at this point that standard and non--standard embedding
vacua fulfil the condition (\ref{constraint}).
%
%
Thus $\epsilon_O=-\epsilon_H$ in (\ref{epsilonO})
implying that the average volume of the
Calabi--Yau space 
turns out to be equal to the
lowest order value
%
$<V>=V$
%
and as a consequence (\ref{gut}) and (\ref{relation2}) simplify.
This will not be the case in the presence of five--branes
as we will see in the next section.

Due also to eq.(\ref{constraint}) the following bounds
\be
-1 < \epsilon_O 
<
1\ ,
\label{epsilonbound}
\ee
must be fulfilled
in order to have positive values for $V_O$ and $V_H$. 
Besides, $\epsilon_O>0$ will imply  
that $V_O$ be larger than $V_H$
and therefore the gauge coupling of the observable
sector 
will be weaker than the gauge coupling of the hidden sector 
(see(\ref{alphaO'})). 
The opposite situation $\epsilon_O<0$ may be obtained
in non--standard embedding models.
$V_O$ is now smaller than
$V_H$ 
and therefore the
gauge coupling of the observable sector 
will be stronger than the one of the 
hidden sector\footnote{In the context of supersymmetry
breaking by gaugino condensation this scenario may have several advantageous
features with respect to scenarios with
$\epsilon_O>0$. For a discussion
about this point see \cite{Lalak}.}.

Notice that using (\ref{kinetic}) one can write
$\epsilon_O$ as
\bea
\epsilon_O 
=
\frac{4-(S+\bar S)}
{(S+\bar S)}\ ,
\label{nueva}
\eea
where $\left(2\pi\alpha_O\right)^{-1}=4$ has been used.
%
Thus with (\ref{epsilonbound}) and (\ref{epsilonO})
one obtains that the dilaton and moduli fields are bounded.
In particular,  
\be
0<\beta_O(T+ \bar T) 
<
2\ , \quad 
2 
<
(S+ \bar S)<4 
\ ,
\label{bounds}
\ee
for $0<\epsilon_O<1$ and
\be
\beta_O(T+ \bar T) 
<
0\ , \quad 
(S+ \bar S)>4 
\ ,
\label{BOUNDSS}
\ee
for $-1<\epsilon_O<0$. 
Note that $\epsilon_O$ 
can approach the limit $-1$ only for very large values of 
($S+\bar S$) and therefore of ($T+\bar T$).

%
%
%
%
%
%
%
%

With all these results we can start now the study of scales and soft
terms in the theory.

\subsection{Scales}

We will discuss first how to obtain $M_{GUT}=3\times 10^{16}$ GeV
in the four--dimensional effective theory from heterotic M--theory
\cite{Witten,Banks-Dine}
taking into account the
higher order corrections studied above to the zeroth--order 
formulae \cite{Mio2}.
On the other hand, 
we will analyze whether the special limits
pointed out in \cite{Benakli2},
in order to lower the scales of the theory, even with the possibility of
obtaining  
an extra dimension as large as a millimetre,
may be
obtained in a natural way \cite{Mio2}.

Let us concentrate first in the case $\beta_O>0$, i.e. in the region 
$0<\epsilon_O<1$ in (\ref{epsilonbound}).
Identifying $M_{GUT}$ with $V_O^{-1/6}$
one obtains from (\ref{relation2}), and  (\ref{relation})
(recall that $<V>=V$):
$M_{11}\approx 6\times 10^{16}$ GeV and
$(\pi\rho)^{-1}\approx (2.5-5.3)\times 10^{15}$ GeV, i.e.
the following pattern $(\pi\rho)^{-1}<V_O^{-1/6}<M_{11}$.
On the other hand, to obtain 
$V_O^{-1/6} =M_{GUT}$ when $\beta_O>0$ 
is quite natural. This can be seen from 
(\ref{gut}) since (\ref{bounds}) implies that $T+\bar T$ and
$S+\bar S$ are essentially of order one.
Let us discuss this point in more detail.
Using (\ref{epsilonO}) and (\ref{nueva}) 
it is interesting to write (\ref{gut}) as
\bea
V_O^{-1/6}
& = &
3.6\times 10^{16} 
\left(\frac{\beta_O}{2 \epsilon_O}\right)^{1/2}
\nonumber\\ 
& \times &
\left(1+\epsilon_O\right)^{5/6}
{\rm GeV}\ .
\label{gut3}
\eea
This is shown in Fig.~\ref{scales1} where
$V_O^{-1/6}$ versus $\epsilon_O$ is plotted.
\EPSFIGURE[ht]{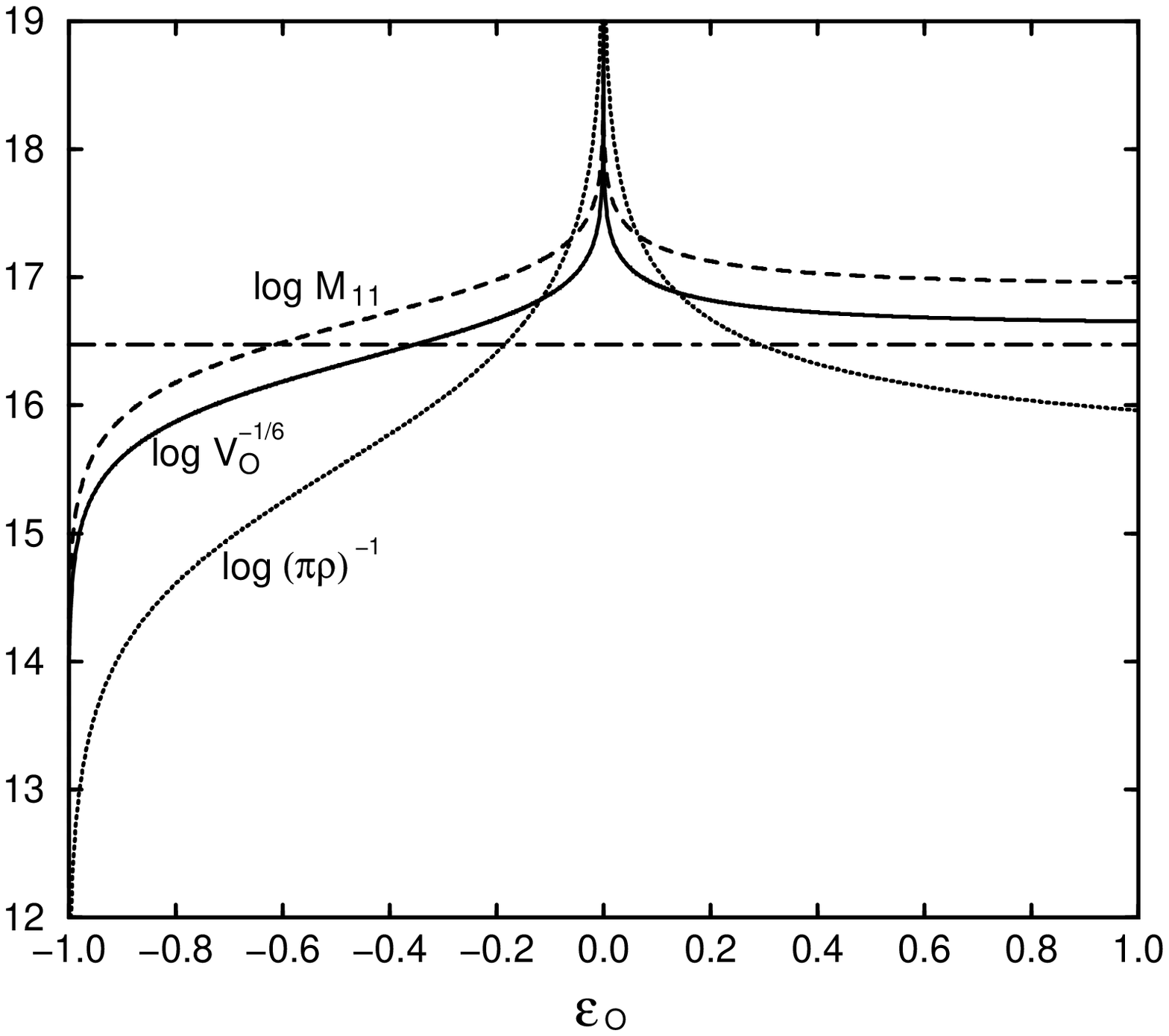,
height=6.0cm,width=6.0cm}
{$\log M_{11}$, $\log V_O^{-1/6}$ and $\log (\pi\rho)^{-1}$ 
versus $\epsilon_O$. The straight line
corresponds to $M_{GUT} = 3\times 10^{16}$ GeV.\label{scales1}}
%
%
The r.h.s. of the figure ($0<\epsilon_O<1$)
corresponds to the case $\beta_O>0$ whereas the l.h.s. 
($-1<\epsilon_O<0$) corresponds to the case  $\beta_O<0$
that will be analyzed below.
For the moment we concentrate on the case $\beta_O>0$ and,
in particular, in Fig.~\ref{scales1} we are showing an example with
$\beta_O=1$.
$(\pi\rho)^{-1}$ and 
$M_{11}$ are also plotted in the figure using (\ref{relation2}) and 
(\ref{relation})
respectively. Most values of $\epsilon_O$ imply 
$V_O^{-1/6} \approx 5\times 10^{16}$ which is quite close 
to the phenomenologically favored value. For example,
for $\epsilon_O=1/4$, which corresponds to $S+\bar S=16/5$ and 
$T+\bar T=4/5$, we obtain 
$V_O^{-1/6} = 6.1\times 10^{16}$ GeV
and for the limit
$\epsilon_O=1$ (as discussed in subsection~2.1, the M--theory
expansion will work even in this limit),
which corresponds to $S+\bar S=T+\bar T=2$, we obtain the lowest
possible
value  
$V_O^{-1/6} = 4.5\times 10^{16}$.

These qualitative results can only be modified in the limit 
$\epsilon_O\rightarrow 0$, i.e. $(T+\bar T)\rightarrow 0$,
since then $V_O^{-1/6}\rightarrow \infty$. Notice that in this case
$(\pi\rho)^{-1}>V_O^{-1/6}$ (see Fig.~\ref{scales1}).
This limit is not interesting not only because $V_O^{-1/6}$ is too
large but also because we are effectively in the weakly--coupled 
region with a very small orbifold radius.

The results for $\beta_O>1$ can easily be deduced from
the figure and eq.(\ref{gut3}).
For those models we are in the limit of
validity if we want to obtain 
$V_O^{-1/6}=M_{GUT}$.
For example, For $\epsilon_O=1$ with $\beta_O=4$, 
$V_O^{-1/6} = 9\times 10^{16}$ GeV.

Let us finally remark that, from the above discussion, it is 
straightforward to deduce that large internal dimensions,
associated with the radius of the Calabi--Yau and/or the radius of 
the orbifold, are not allowed.

Let us now study the value of the scales in models with 
$\beta_O<0$. We can use again
(\ref{gut3}), but now with $-1<\epsilon_O<0$.
This is shown in the l.h.s.  
of the Fig.~\ref{scales1}. 
Unlike the previous models where always 
$V_O^{-1/6}$ was bigger than $M_{GUT}$
for any $\beta_O>0$, in these non--standard
embedding models $M_{GUT}$ can be obtained. For example in the
case shown in the figure, $\beta_O=-1$, with 
$\epsilon_O=-0.35$
which, using (\ref{nueva}) and (\ref{epsilonO}), 
corresponds to $S+\bar S=6.15$ and 
$T+\bar T=2.15$, we obtain
$V_O^{-1/6} = 3\times 10^{16}$ GeV.
For other values of $\beta_O$ this is also possible. Notice that the 
figure for $V_O^{-1/6}$ will be the same  adding the constant
$\log |\beta_O|^{1/2}$ and therefore there will be lines, corresponding
to $V_O^{-1/6}$, intersecting with the
straight
line corresponding to  
$M_{GUT}$.
In this sense, if we want to obtain models with the
phenomenologically favored GUT scale, non--standard embedding
models with $\beta_O<0$
are more compelling than models with $\beta_O>0$.

On the other hand, 
for $\beta_O>0$
we obtained above the lower bound $\approx 10^{16}$ GeV for all scales of
the theory (see the r.h.s. of Fig.~\ref{scales1}),
far away from any direct experimental detection. Now we want
to study this issue in cases with $\beta_O<0$.
From (\ref{gut3}), clearly
in the limit $\epsilon_O\rightarrow -1$ we are able to obtain
$V_O^{-1/6}\rightarrow 0$ and therefore, given (\ref{relation2}) also
$(\pi\rho)^{-1}\rightarrow 0$ (see the l.h.s.of Fig.~\ref{scales1}).  
Thus to lower the scale $V_O^{-1/6}$ down to the experimental bound
(due to Kaluza--Klein excitations) of $1$ TeV 
is possible in this limit. However, this is true only
for values of $\epsilon_O$ extremely close to $-1$.
For example, 
for $\epsilon_O=-0.999999$ 
which, using (\ref{nueva}) and (\ref{epsilonO}),
corresponds to $S+\bar S=4\times 10^{6}$ and 
$T+\bar T=4\times 10^{6}-4$, we obtain the intermediate scale 
$V_O^{-1/6} = 2.5\times 10^{11}$ GeV, i.e. $M_{11}=5\times 10^{11}$ GeV,
with
$(\pi\rho)^{-1} = 3\times 10^{6}$ GeV. This is an interesting
possibility
since 
an intermediate scale $\approx 10^{11}$ GeV was proposed in
\cite{Benakli2} in order to solve some phenomenological problems and in
\cite{Nosotros} in order to solve the $M_W/M_{Planck}$ 
hierarchy problem\footnote{
For example, for a D3--brane in type I strings where
$\frac{M_W}{M_{Planck}}\approx 
\frac{\alpha_O}{2}\left(\frac{M_c}{M_I}\right)^6$, 
with a modest
input hierarchy between string and compactification scales,
$M_I\approx 10^{11}$ GeV and $M_c\approx 10^9$ GeV,  
one obtains the desired hierarchy 
$M_W/M_{Planck}\approx 10^{-16}$ without invoking any hierarchically
suppressed
nonperturbative effect like e.g. gaugino condensation.
However, it is worth noticing that those values would imply
${\rm Re}(S)=1/\alpha_O\approx 24$ and
${\rm Re}(T)=\frac{1}{\alpha_O}\left(\frac{M_I}{M_c}\right)^4 \approx
10^9$,
i.e. one has again a hierarchy problem but now for the vev of the 
fields that one has to determine dynamically.}. 
In any case, 
it is obvious that the smaller the scale the larger the amount of 
fine--tuning becomes. The experimental lower bound for the scale $V_O^{-1/6}$,
$1$ TeV, can be obtained with
$\epsilon_O=10^{-16}-1$, i.e. 
$S+\bar S=4\times 10^{16}$ and 
$T+\bar T=4\times 10^{16}-4$. Then one gets
$V_O^{-1/6}=1181.5$ GeV with $(\pi\rho)^{-1} = 3.2\times 10^{-9}$ GeV.
Since only gravity is free to propagate in the orbifold, this extremely
small value is not a problem from the experimental point of view.
In any case,
it is clear that low scales are possible
but the fine--tuning needed renders the situation highly
unnatural. Another problem related with the limit 
$\epsilon_O\rightarrow -1$ will be found below when studying soft
terms,
since $|M|/m_{3/2}\rightarrow \infty$. Thus a extremely small gravitino
mass is needed to fine tune the gaugino mass $M$ to the $1$ TeV scale 
in order to avoid the gauge hierarchy problem.

There is a value of $\beta_O$ which is in principle allowed and has not been
analyzed yet. This is the case $\beta_O=0$. As we will see in a moment,
to lower the scales a lot in this context is again possible.
Since $\epsilon_O$ in (\ref{epsilonO}) is vanishing and using 
(\ref{nueva}),
$S+\bar S=4$, eq. (\ref{gut}) can be written as 
\be
V_O^{-1/6}
=
3.6\times 10^{16} 
\left(\frac{2}{T+ \bar T}\right)^{1/2}  
{\rm GeV}
\ .
\label{ggut2}
\ee
This is plotted in Fig.~\ref{scales2} together with 
$(\pi\rho)^{-1}$ and $M_{11}$. We see that the value 
$V_O^{-1/6} = 3\times 10^{16}$ GeV is obtained
for the reasonable value $T+\bar T=2.88$. 
\EPSFIGURE[ht]{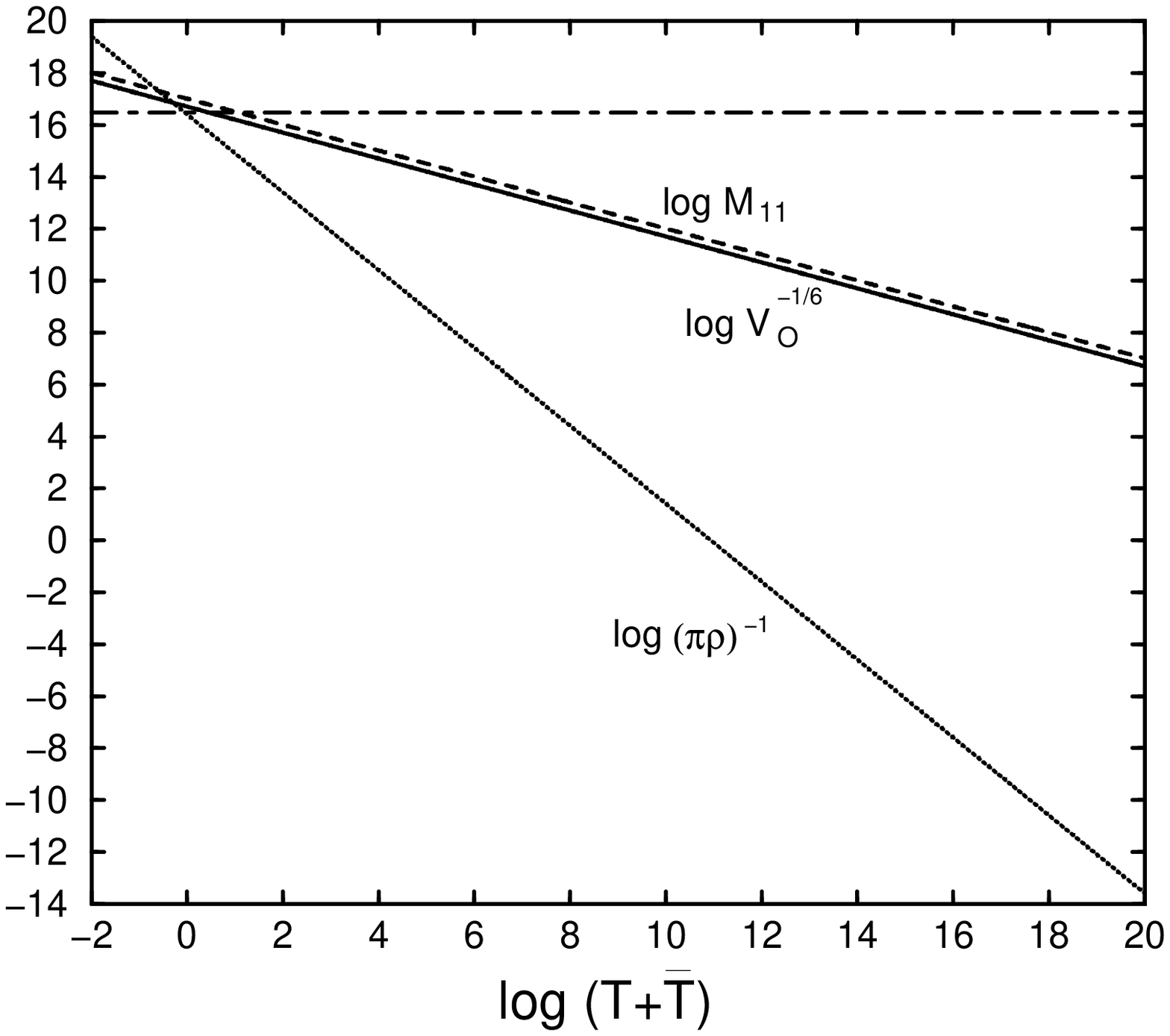,height=6cm,width=6cm}
{$\log M_{11}$, $\log V_O^{-1/6}$ and $\log (\pi\rho)^{-1}$ 
versus $\epsilon_O$. The straight line
corresponds to $M_{GUT} = 3\times 10^{16}$ GeV.\label{scales2}}
On the other hand, 
the larger $T+\bar T$ the smaller 
$V_O^{-1/6}$ becomes. 
The lower bound for $V_O^{-1/6}$ is obtained 
with $T+\bar T=4\times 10^{19}$ GeV. Then
one gets $V_O^{-1/6}=8\times 10^{6}$ GeV and
$(\pi\rho)^{-1}=10^{-13}$ GeV. Smaller values of  
$V_O^{-1/6}$ are not allowed since experimental results on the
force of gravity constrain 
$(\pi\rho)$ to be less than a millimetre.
Thus, although very low scales are allowed for the particular value
$\beta_O=0$, clearly we introduce a hierarchy problem between
$S+\bar S$ and $T+\bar T$.

\subsection{Soft terms}
\EPSFIGURE[ht]{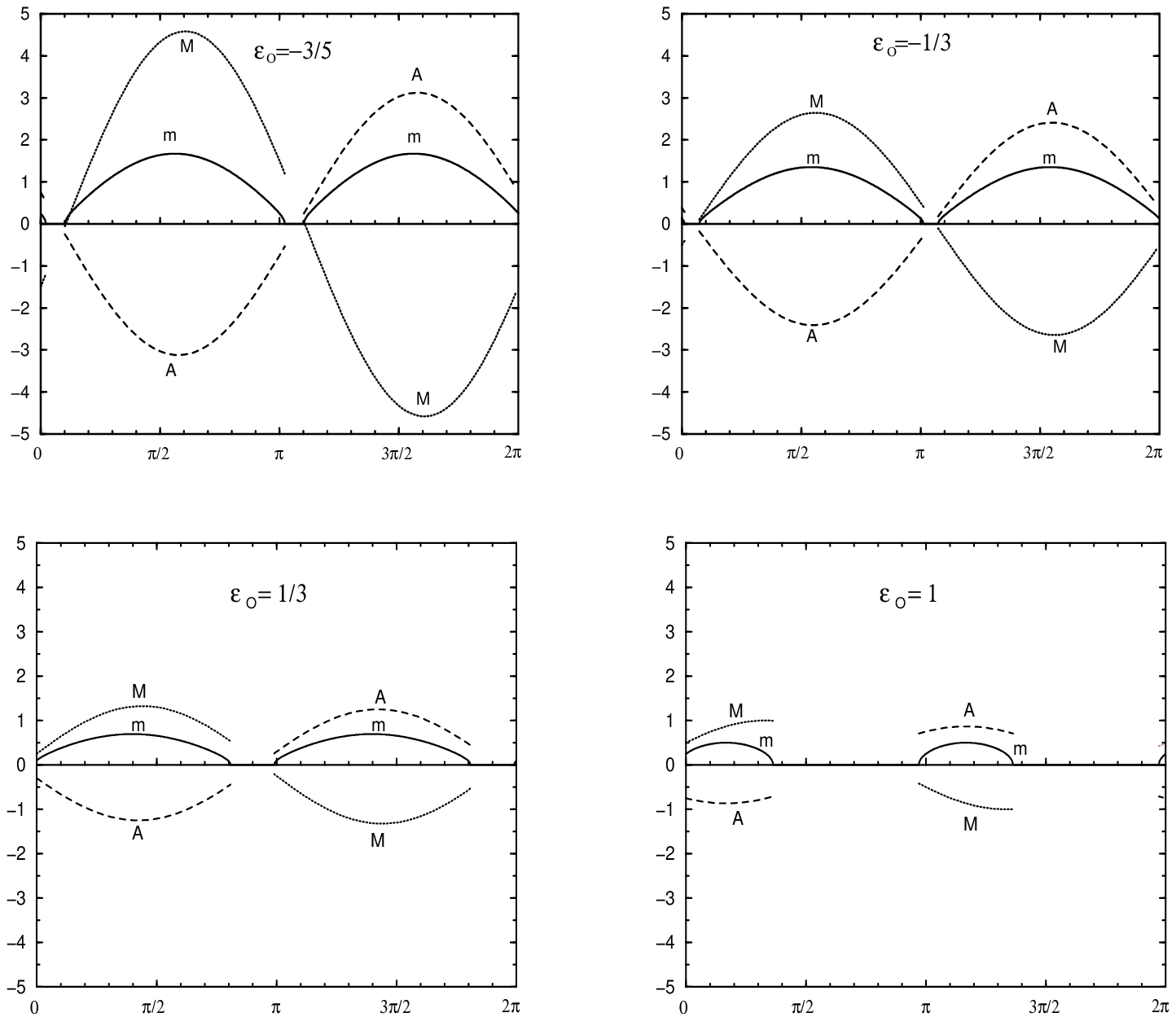,
width=12.4cm, height=8.0cm}
{Soft parameters in units of $m_{3/2}$ versus $\theta$ for
different values of $\epsilon_O$. Here
$M$, $m$
and $A$ are the gaugino mass, the scalar mass and the trilinear parameter
respectively.\label{nonstandardsoft}}
Applying the standard
(tree level) soft term formulae
\cite{
Brignole-Ibanez-Munoz} 
for the above supergravity model given by
(\ref{kahler}), (\ref{kinetic}) and (\ref{superpotential}), one can
compute the soft terms 
straightforwardly\footnote{
Unlike \cite{Lukas-Ovrut-Waldram2} where only linear terms in $\epsilon_1$
are kept, we keep all contributions to soft terms
avoiding accidental cancellations at linear order, e.g. in scalar masses.
Higher order
terms in (\ref{kahler}) might modify the higher order contributions
but, as argued in subsection 2.1 these terms will be suppressed.} \cite{Mio}
%
\bea
M &=& \frac{\sqrt{3}m_{3/2}}{1+\epsilon_O}\left(\sin\theta + 
\frac{1}{\sqrt{3}}\epsilon_O\cos\theta \right)
\ ,
\nonumber \\
m^2 &=& m_{3/2}^2 
- 
\frac{3m_{3/2}^2}{\left(3+\epsilon_O\right)^2}\left[
\epsilon_O\left(6+\epsilon_O\right)sin^2\theta\right. 
\nonumber\\ 
&+& 
\left.\left(3+2\epsilon_O\right)\cos^2\theta
- 2\sqrt{3}\epsilon_O\sin\theta\cos\theta \right]\ , 
\nonumber \\
A &=&
-\frac{\sqrt{3}m_{3/2}}{3+\epsilon_O}\left[\left(3-2\epsilon_O\right)
\sin\theta 
+ \sqrt{3}\epsilon_O\cos\theta \right] \ ,
\nonumber\\
\label{softterms}
\eea
where $m_{3/2}$ is the gravitino mass, and
vanishing cosmological constant and phases are assumed,
given the current experimental limits.
Here $M$, $m$ and $A$ denote gaugino masses, scalar masses and 
trilinear
parameters respectively. The bilinear $B$ parameter
can be found in \cite{Bailin,Kokorelis,Mio2}.
We are using here the parameterization 
introduced in \cite{Brignole-Ibanez-Munoz2}
in order to know what fields, either $S$ or $T$, play the predominant role
in the process of supersymmetry breaking
$F^S = \sqrt 3 m_{3/2}(S+\bar S)\sin\theta$ , 
$F^T = m_{3/2}(T+\bar T)\cos\theta$.

%

As mentioned in the introduction, the structure of these soft terms is
qualitatively different from
that of a Calabi--Yau compactification of the (tree--level) 
weakly--coupled heterotic string
found in \cite{Brignole-Ibanez-Munoz2} which can be recovered
from (\ref{softterms}) by 
taking the limit
$(T+\bar T)\ll (S+\bar S)$, i.e. $\epsilon_O\rightarrow 0$:
\bea
-A = M = \sqrt 3 m = \sqrt 3 m_{3/2} \sin\theta\ . 
\label{wsoft}
\eea
%
%
Clearly the M--theory result (\ref{softterms}) is more involved
due to the additional dependence on $\epsilon_O$.
Nevertheless we can simplify the analysis by taking into account
the bounds (\ref{epsilonbound}).

We show in Fig.~\ref{nonstandardsoft} the dependence on $\theta$
of
the soft terms $M$, $m$, and $A$ in units of the gravitino mass for
different values of $\epsilon_O$ \cite{Mio,Lii,Mio2}.
Several comments are in order.
First of all, some ranges of $\theta$ are forbidden
by having a  negative scalar mass-squared.
In the weakly--coupled heterotic string case
shown in Fig.~\ref{choi}, the forbidden region vanishes since
the squared scalar masses are always positive (see (\ref{wsoft})).
About the possible range of soft terms,
the smaller the value of $\epsilon_O$, the larger the range becomes.
In the limit
$\epsilon_O\rightarrow -1$, $0.3<|A|/m_{3/2}<4.58$,  
$0<m/m_{3/2}<2.26$ and $|M| \rightarrow \infty$.

In order to discuss the supersymmetric spectra further, it is worth
noticing that gaugino mas\-ses are in general larger than 
scalar masses.
This implies at low--energy  
($\approx M_W$)
the following qualitative result \cite{Mio}: 
$M_{\tilde g}\approx m_{\tilde q}>m_{\tilde l}$,
where $\tilde g$ denote the gluino, 
$\tilde l$ all the sleptons and $\tilde q$
first and second generation squarks. 
Other analyses taking into account the details of the electroweak radiative
breaking can be found in \cite{Bailin,Kawamura}.
Only 
for values of $\epsilon_O$ approaching $-1$ 
the opposite situation, scalars heavier than gauginos, 
may occur. This is for two narrow ranges of values of $\theta$
as can be seen in Fig.~\ref{nonstandardsoft} for $\epsilon_O=-3/5$.
Let us remark that $M/m_{3/2}$ and $m/m_{3/2}$ are then very
small
and therefore $m_{3/2}$ must be large in order to fulfil e.g. the
low--energy
bounds on gluino masses. In this special limits $m\gg M$ is possible
and then $M_{\tilde g}<m_{\tilde q}\approx m_{\tilde l}$ \cite{Mio2}.

Notice that in the (tree--level) weakly--coupled heterotic string,
the limit $\sin\theta\rightarrow 0$ is not well defined since
all $M$, $A$, $m$ vanish in that limit. One then has to include
the string one--loop corrections (or the sigma--model corrections)
to the K\"ahler potential and gauge kinetic functions which would
modify the boundary conditions (\ref{wsoft}). This is similar to
what happens in orbifold compactifications where, at the end of the
day, scalars are heavier than gauginos due to string loop
corrections \cite{Brignole-Ibanez-Munoz2}.
This problem is not
present in the heterotic M--theory, as can be deduced from
Fig.~\ref{nonstandardsoft}, except in models with
$\beta_O=0$, i.e. $\epsilon_O=0$ and therefore with 
boundary conditions (\ref{wsoft}). 

\EPSFIGURE[ht]{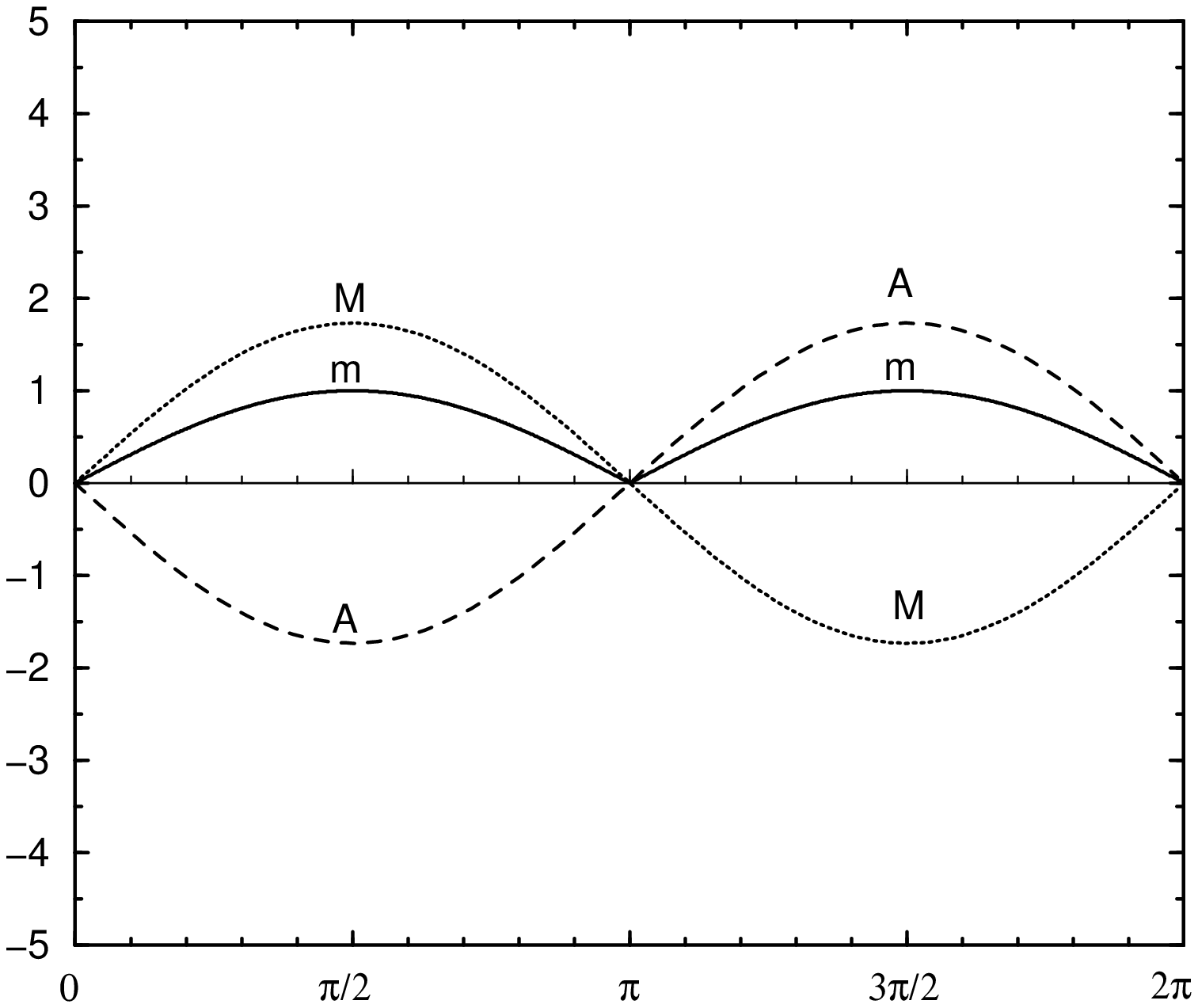,height=3.5cm,width=5.0cm}
{The same as Fig.~\ref{nonstandardsoft}
but for the weakly--coupled heterotic string limit.\label{choi}}

\subsection{Charge and colour breaking}

\EPSFIGURE[hb]{fig7abis.ps,height=5.5cm,width=6.0cm}
{Excluded regions of the parameter space.\label{ccbM}}

We discussed in subsection~2.2 how effective supergravity models 
from heterotic M--theory can
be strongly constrained by imposing the (experimental) requirement
of universal soft scalar mass\-es, 
in order to avoid dangerous 
flavour changing neutral current phenomena. We can go further and
impose the (theoretical)
constraint of demanding the no existence of low--energy charge and
colour breaking minima deeper than the standard vacuum \cite{CCB}.
In this type of analysis, the form of the soft terms is crucial.
In the case of the standard embedding, $0<\epsilon_O<1$, with soft
terms
given by (\ref{softterms})
the restrictions are very strong and the whole parameter space
($m_{3/2},\theta ,\epsilon_O ,B$) turns out to be excluded on these 
grounds \cite{Savoy,Alejandro}. 
This is shown in Fig.~\ref{ccbM} \cite{Alejandro}
for a fixed value of $m_{3/2}$ (or, equivalently, of $m$) with
$\epsilon_O=1$.
Then we are left with two independent parameters $B$ and $\theta$.
Whereas the black region is excluded because it is not possible to reproduce
the experimental mass of the top, the rest is excluded
by charge and colour breaking constraints. The small squares indicate regions
excluded by the so--called UFB constraints and
the circles indicate regions excluded by the so--called CCB constraints.
Other values for $m$ and $\epsilon_O$ do not modify these conclusions.

Given these dramatic consequences,
a way--out must be searched\footnote{We could  
accept that we live in a metastable vacuum, provided its lifetime
is longer than the present age of the Universe, thus rescuing points
in the parameter space \cite{CCB}.
In this sense the constraints found are basically the most conservative
ones (in the sense of safe ones).} \cite{prepa}.
The first possibility is 
to consider the case of the non--standard embedding
since although the formulae for the soft terms are the same
(\ref{softterms})
the parameter space is different:
$-1<\epsilon_O<0$.
Another possible way--out is to consider the presence of five--branes
in the vacuum, then the soft terms  are different (see(\ref{una}))
and new parameters, as e.g. the goldstino angles $\theta_n$ associated with
$F$--terms
of the five--branes,
enter in the game. Possibly some regions in the parameter space will
be allowed. Although now the situation is clearly more model dependent.

It is worth noticing that the situation in the perturbative 
heterotic string compactified on a Calabi--Yau is basically worst.
There the whole parameter space ($m_{3/2},\theta ,B$)
is forbidden \cite{Alejandro} 
and there is no the freedom of playing around with
$\epsilon_O$ and/or $\theta_n$ from five--branes.
Only in the limit $\sin\theta\rightarrow 0$, where one has to include
loop corrections to the boundary conditions (\ref{wsoft}), small
regions
might be allowed. At least this is the case of orbifold
compactifications with the same boundary conditions (i.e. models
where all observable particles have modular weight $-1$) \cite{Alejandro}.

\section{Vacua with five-branes}

In the previous section, we studied the phenome\-nology of heterotic
M--theory vacua obtained thro\-ugh standard and non--standard
embedding.
Here we want to analyze (non--perturbative) heterotic M--theory vacua
due to the presence of five--branes. 

\subsection{Four--dimensional effective supergravity}

As mentioned in the introduction, five--branes are non--perturbative objects, 
located at points, $x^{11}=x_n (n=1,...,N)$, throughout the orbifold interval.
The modifications to the four--dimensional effective action determined by
(\ref{kahler}), (\ref{kinetic}) and (\ref{superpotential}), 
due to their presence, have recently been investigated by
Lukas, Ovrut and Waldram \cite{five-branes,five-branes2}. Basically,
they are due to the existence of moduli, $Z_n$, whose
${\rm Re}(Z_n)\equiv z_n=x_n/\pi\rho \in (0,1)$ 
are the five--brane positions in the normalized orbifold coordinates.
Then, the effective supergravity obtained from heterotic M--theory
compactified 
on a Calabi--Yau manifold in the presence of five--branes is now determined by
\bea
K &=& -\ln (S+\bar{S}) -3\ln (T+\bar{T})+ K_5 
\nonumber\\ 
& + & \frac{3}{T+\bar{T}}
\left(1+\frac{1}{3}
e_O\right) H_{pq}C_O^p \bar{C}_O^q,
\nonumber\\
f_{O} &=& S+
B_O
T\ , \quad f_{H}=S+
B_H
T\ ,\nonumber \\
W_O &=& d_{pqr}C_O^pC_O^qC_O^r\ ,
\label{kahlerbranas}
\eea
with $e_O=b_O  \frac{T+\bar T}{S+\bar S}$.
Here
$K_5$ is the K\"ahler potential for the five--brane moduli
$Z_n$, $H_{pq}$ is some $T$--independent metric (see footnote~4) and
$b_O = \beta_O + \sum_{n=1}^N(1-z_n)^2\beta_n$,
$B_O = \beta_O + \sum_{n=1}^N(1-Z_n)^2\beta_n$,
$B_H = \beta_H + \sum_{n=1}^N(Z_n)^2\beta_n$,
%
with $\beta_O$, $\beta_H$ the instanton numbers and $\beta_n$
the five--brane charges. The former, instead of condition (\ref{constraint}),
must fulfil now:
$\beta_O + \sum_{n=1}^N\beta_n +\beta_H=0$.
%

\subsection{Phenomenology}

Assuming for simplicity that $<Z_n>=<z_n>$, i.e. $<B_O>=<b_O>$, 
(\ref{alphaO'}) 
is still valid with the
modification
$\epsilon_{O,H}\rightarrow e_{O,H}$, where
$e_H=b_H\frac{T+\bar T}{S+\bar S}$
with 
$b_H = \beta_H + \sum_{n=1}^N(z_n)^2\beta_n$.
%
Following the
analysis
of section 3 
one can write $e_O$ as a function of $S+\bar S$ 
as in (\ref{nueva})
%
%
and therefore the bounds for $S+\bar S$ in (\ref{bounds}) and (\ref{BOUNDSS})
are still valid if $-1<e_O<1$ is possible.
In fact one can obtain different bounds on $e_O$ 
depending
on the sign of both $b_O$ and $b_H$ \cite{Mio2}. For example,
if $b_H\geq 0$ and  $b_O\leq 0$, then
$e_H$ is positive and $e_O$ negative. Since
$V_O=V(1+e_O)$ must be positive we need
%
$-1<e_O\leq 0$.
Another example is the case $b_H\geq 0$ and $b_O>0$.
Now since $e_O$ is positive $V_O$ will always be positive and 
therefore
the only bound is 
$e_O>0$.
It is worth noticing that the values $0<(S+\bar S)<2$, corresponding to 
$e_O>1$ are then possible. This was not the case in the
absence of five--branes since $\epsilon_O>1$ was not allowed.

\subsubsection{Scales}

In the presence of five--branes 
$V=<V>$ as in section~3
is no longer true since
$V_{O,H}=V(1+e_{O,H})$ with $e_O+e_H\neq0$ in general.
Therefore 
$\frac{V}{<V>}=
\left[1+\frac{e_O}{2}\left(1+\frac{b_H}{b_O}\right)\right]^{-1}$ and
%
the relevant formulae to study the relation between the different
scales of the theory are (\ref{gut})  and
(\ref{relation2}) with the modification $\epsilon_O\rightarrow e_O$.
Notice that (\ref{relation}) is not modified.
Similarly to the case without five--branes, to obtain
$V_O^{-1/6}\approx 3\times 10^{16}$ GeV when
$T+\bar T$ and $S+\bar S$ are of order one is quite natural. 
%
%
To carry out the numerical analysis we can use (\ref{gut3}) 
with the factor $\frac{V}{<V>}$ written above
multiplying the r.h.s.
and with
the modifications $\epsilon_O\rightarrow e_O$, $\beta_O\rightarrow b_O$.
%
%
Several examples were considered in \cite{Mio2}.
Although the qualitative results are similar to those of
Fig.~\ref{scales1} with $e_O$ instead of $\epsilon_O$, 
now the line corresponding to  $V_O^{-1/6}$ in the r.h.s. of the 
figure may intersect
the straight line corresponding to the GUT scale.
Of course this effect, which is due essentially 
to the extra factor discussed above, is welcome.
%
\EPSFIGURE[ht]{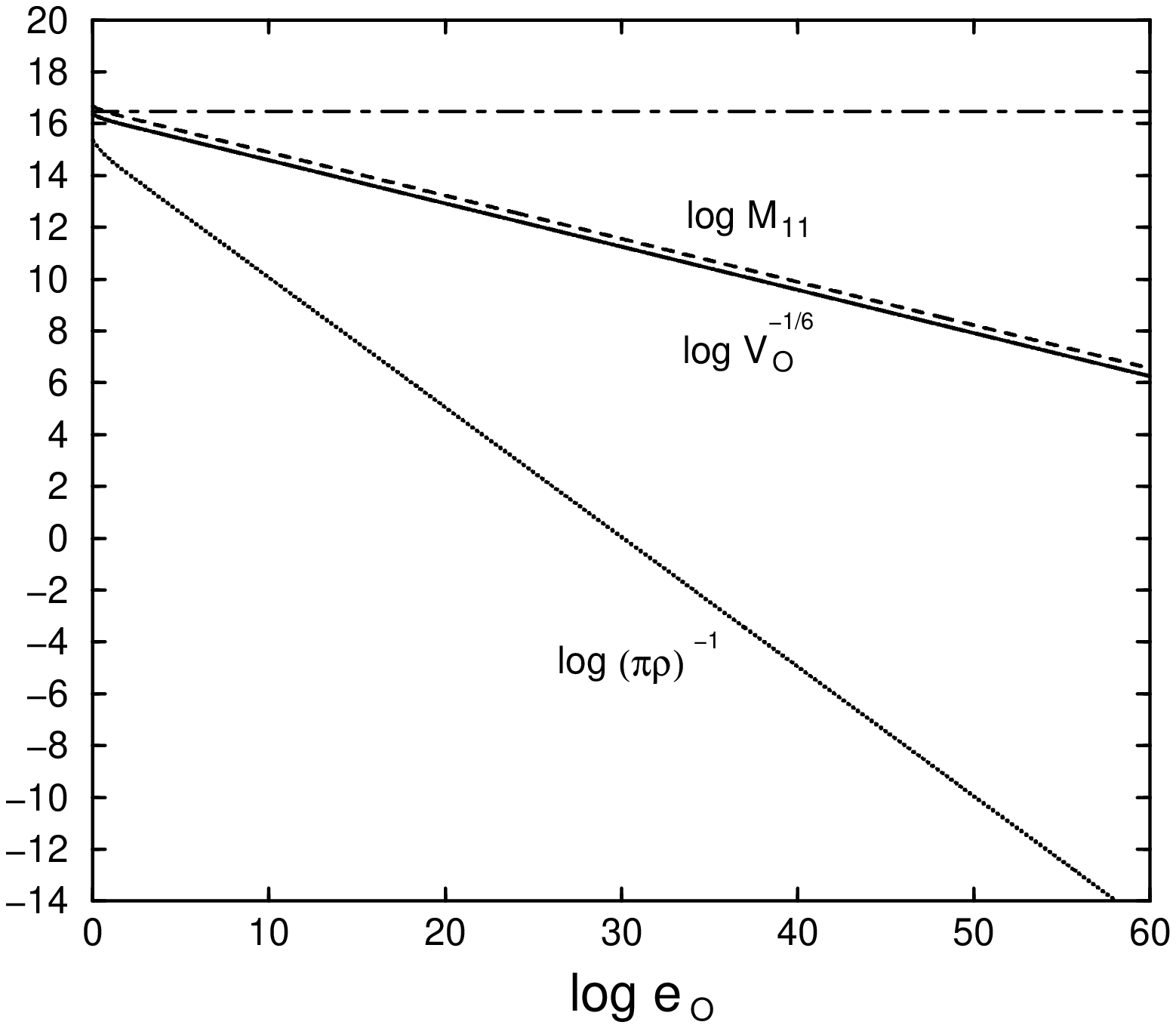,height=5.0cm,width=5.0cm}
{$\log M_{11}$, $\log V_O^{-1/6}$ and $\log (\pi\rho)^{-1}$ 
versus $\log e_O$. The straight line
corresponds to $M_{GUT} = 3\times 10^{16}$ GeV.\label{scales1five}}
%

Only in some special limits one may lower the scales. 
As in the case without five--branes, fine--tuning
$e_O\rightarrow -1$
we are able to obtain  $V_O^{-1/6}$ as low as we wish.
The numerical results will be basically similar 
to the ones of 
subsection 3.1.
Moreover, as discussed above, 
$e_O>1$ is possible in the presence of five--branes
and therefore with $e_O$ sufficiently large we may get $V_O^{-1/6}$ 
very small. This is shown in Fig.~\ref{scales1five} for the example
$b_O=b_H=1/2$.
For instance, with $\log e_O=56.1$ the experimental lower bound 
$(\pi\rho)^{-1}=10^{-13}$ GeV
is obtained for  $V_O^{-1/6}=8\times 10^{6}$ GeV, corresponding  
to $S+\bar S=3.1\times 10^{-56}$ and 
$T+\bar T=8$.
Clearly we introduce a hierarchy problem.

%

Finally, the analysis of
the special case $b_O=0$ will be
similar
to the one of the case $\beta_O=0$ without five--branes in subsection
3.1.
We can use (\ref{ggut2}) 
with the average volume
$[1+b_H(T+\bar T)/8]^{-1/2}$ multiplying the r.h.s.
%
%
Depending on the value of $b_H$
we obtain different results \cite{Mio2}. 
For example if $b_H>0$ the results are qualitatively similar
to those of Fig.~\ref{scales2}, the larger
$T+\bar T$ the smaller $V_O^{-1/6}$ becomes. However, notice that now
for large $T$ 
we have a factor $(T+\bar T)^{-1}$ and then not so large values of 
$T+\bar T$  as in Fig.~\ref{scales2} are needed in order to lower the scales.
For example, if $b_H=1$
then $V_O^{-1/6}=1$ TeV can be obtained for $T+\bar T=10^{14}$ 
with the size of
the orbifold $(\pi\rho)^{-1}=5\times 10^{-12}$ GeV close to  its experimental
bound of $1$ millimetre. 
In any case, still a large hierarchy between $S+\bar S$ and
$T+\bar T$ is needed.

\subsubsection{Soft terms}

Let us now concentrate on the computation of soft terms
\cite{five-branes2,Kubo,Mio2}. 
Due to the possible contribution of several $F$--terms associated with 
five--branes, which can have in principle off--diagonal
K\"ahler
metrics, the computation of the soft terms turns out to be
extremely
involved. In order to get an idea of their value and also to study
the deviations with respect to the case without five--branes we can
do some simplifications. One possibility is to assume that 
five--branes are present but only the
$F$--terms
associated with the dilaton and the modulus contribute to supersymmetry
breaking,
i.e. $F^{Z_n}=0$. Then, assuming as before $<Z_n>=<z_n>$,
eq.(\ref{softterms}) is still valid with 
$e_O$ instead of $\epsilon_O$.
Under these simplifying assumptions, Fig.~\ref{nonstandardsoft} 
is also valid in this case 
since, as discussed above,
the
range of allowed values of $e_O$ includes those of $\epsilon_O$, 
i.e. $-1<e_O<1$. The relevant difference with respect to the case
without
five--branes is that now values with $e_O\geq 1$ are allowed.
This possibility was studied 
in \cite{Mio2}.
Although 
the soft terms are qualitatively different from those without
five--branes analyzed in Fig.~\ref{nonstandardsoft},
the fact that  always scalar masses are smaller than gaugino masses
is still true for $e_O\geq 1$. 
As discussed below (\ref{wsoft}),
we will obtain at low--energies,
$M_{\tilde g}\approx m_{\tilde q}>m_{\tilde l}$.
%
%
%
%
%

Another possibility to simplify the computation of the soft 
terms is to assume that there is only one five--brane in the model. 
For example,
parameterizing $F^S = \sqrt{3} m_{3/2}  (S+\bar S)\sin\theta\cos\theta_1$,
$F^T =  m_{3/2}  (T+\bar T)\cos\theta\cos\theta_1$, 
$F^{Z_1} = \sqrt{3}  m_{3/2}\\
({{\partial}_1{\partial}_{\bar 1 }K_5})^{-1/2}sin\theta_1$,
%
where $\theta_1$ is the new goldstino angle associated to the $F$--term
of the five--brane, 
one obtains \cite{Kubo,Mio2}.  
%
\bea
M&=&\frac{ \sqrt{3} m_{3/2}}
{\left(1+\frac{B_O T + \bar B_O \bar T}{S+ \bar S}\right)}
\left( \sin\theta\cos\theta_1\right.
\nonumber\\ 
&+&
\frac{1}{\sqrt{3}}B_O  \frac{e_O}{b_O} 
\cos\theta\cos\theta_1 
\nonumber \\
&-&
\left. \frac{2T}{S+\bar S}(1-Z_1)\beta_1 
({{\partial}_1{\partial}_{\bar 1 }K_5})^{-1/2}
sin\theta_1  \right)  
\ ,
\nonumber \\
m^2&=&  m_{3/2}^2-\frac{3 m_{3/2}^2}{(3+e_O)^2}\left\{ e_O 
(6+e_O)\sin^2\theta  \right.
\nonumber \\
&\times& \cos^2\theta_1 +(3+2e_O)\cos^2\theta\cos^2\theta_1  
\nonumber\\ 
&-& 2\sqrt{3}e_O\sin\theta\cos\theta\cos^2\theta_1 
\nonumber\\
&+& ({{\partial}_1{\partial}_{\bar 1 }K_5})^{-1}
\sin^2\theta_1 \left( (3+e_O)  
\beta_1\frac{e_O}{2b_O}\right.
\nonumber\\
&-& \left. \left[(1-z_1) 
\beta_1 \frac{e_O}{b_O}\right]^2 \right)
+ 6(1-z_1)\beta_1\frac{e_O}{b_O}
\nonumber\\ 
&\times& ({{\partial}_1{\partial}_{\bar 1 }K_5})^{-1/2}
\sin\theta\sin\theta_1\cos\theta_1
\nonumber \\
&-&  2\sqrt{3}(1-z_1)\beta_1\frac{e_O}{b_O} 
({{\partial}_1{\partial}_{\bar 1 }K_5})^{-1/2}
\nonumber\\ 
& \times & \left.\cos\theta\sin\theta_1\cos\theta_1\right\} \ , 
\nonumber\\
A&=&-\frac{\sqrt{3} m_{3/2}C}{3+e_O}\left\{ (3-2e_O) 
\sin\theta\cos\theta_1 \right.
\nonumber\\ 
&+& \sqrt{3}e_O\cos\theta\cos\theta_1
-
({{\partial}_1{\partial}_{\bar 1 }K_5})^{-1/2}
\sin\theta_1
\nonumber\\
&\times& \left. 
\left[(3+e_O)
{{\partial}_1 K_5}
+ 3(1-z_1)\beta_1 \frac{e_O}{b_O} \right] \right\}
\label{una}
\eea
The formula for the $B$ parameter can be found in \cite{Mio2}.
Unfortunately, the numerical analysis of this simplified case is not
straightforward. All soft terms depend not only on the new goldstino
angle $\theta_1$ in addition to 
$m_{3/2}$, $\theta$
and $e_O$, but also on other free parameters.
For example, although gaugino masses can be further 
simplified with the assumption
$<Z_n>=<z_n>$, i.e. $<B_O>=<\bar B_O>=<b_O>$ 
(and therefore $(B_O T + \bar B_O \bar T)/(S+\bar S)=e_O$), 
and $<T>=<\bar T>$ 
(and therefore $2T/(S+\bar S)=e_O/b_O$),
%
%
still
they have an explicit dependence on 
$z_1$ and 
$\partial_1 \partial_{\bar 1} K_5$.
Notice that, for a given model,
$\beta_O$ and $\beta_1$ are known and therefore $b_O$ can be computed
once $z_1$ is fixed.
Something similar occurs for the 
$A$ parameter, where $z_1$, $\partial_1 K_5$ and
$\partial_1 \partial_{\bar 1} K_5$ appear explicitly, and
for the scalar masses, where 
 $z_1$ and
$\partial_1 \partial_{\bar 1} K_5$ also appear.
Thus in order to compute soft terms when a five--brane is present and
contributing to supersymmetry breaking we have to input these values.
Fortunately, $z_1$ is in the range $0<z_1<1$ and, although
$K_5$ is not known,
since
it depends on $z_1$,
we expect 
$\partial_1 K_5$, $\partial_1 \partial_{\bar 1} K_5 = {\cal O}(1)$.
So we can consider the following representative case:
$z_1=1/2$ and $\partial_1 K_5=\partial_1\partial_{\bar 1}K_5=1$.
Since, still we have to input the value of $b_O$, we choose an
example with $\beta_O=1$ and $\beta_1=-2$ which implies 
$b_O=b_H=1/2$.
Then
all positive values of $e_O$ are allowed. 
We 
show in Fig.~\ref{susyfivesoft+} the soft terms for the value
$e_O=3/5$ with $\theta_1=\pi/3$.
\EPSFIGURE[ht]{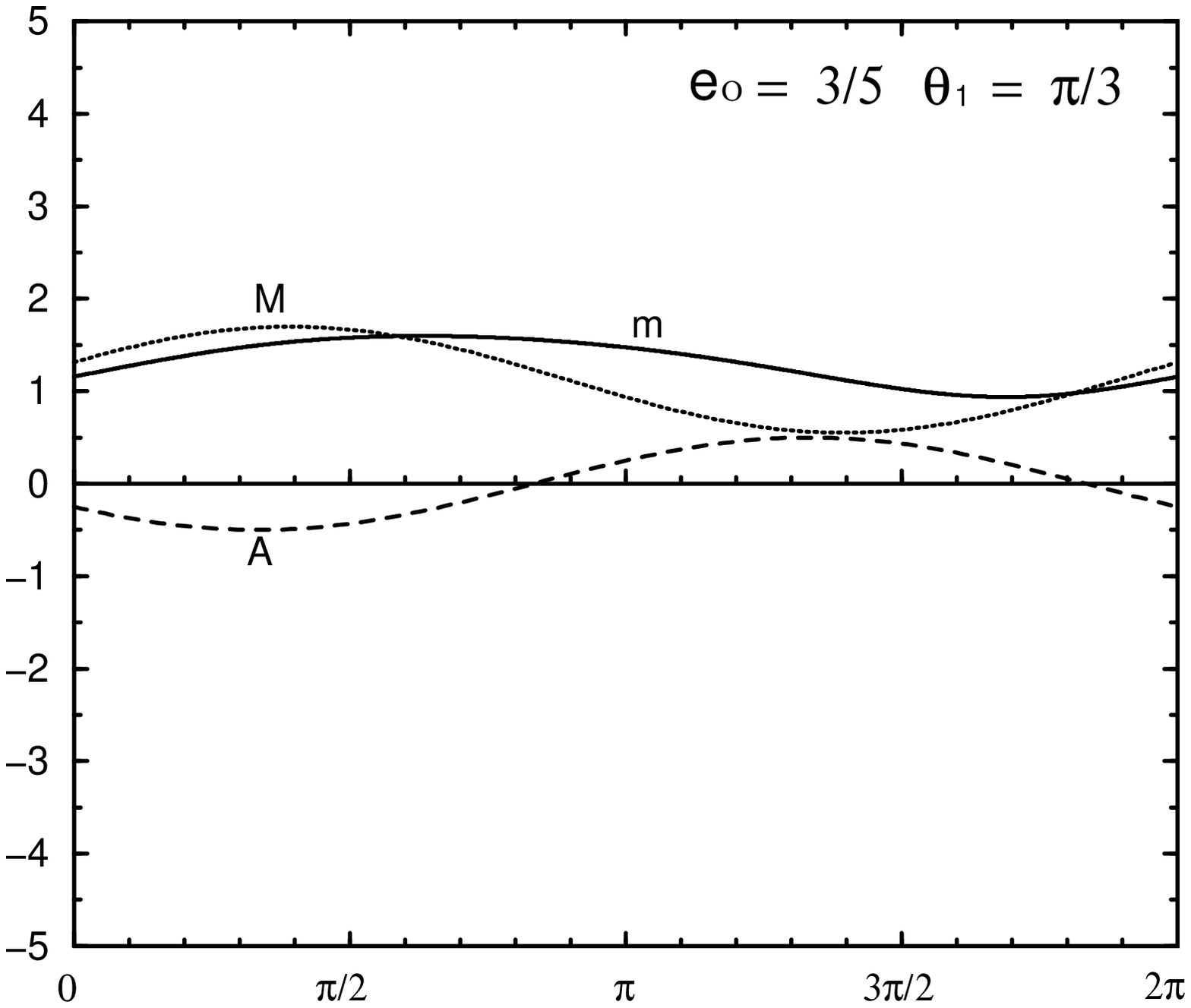,width=6.2cm, height=4.3cm}
{Soft parameters in units of $m_{3/2}$ versus $\theta$ 
when one five--brane is
contributing to supersymmetry breaking.\label{susyfivesoft+}}
Unlike Fig.~\ref{nonstandardsoft} without five--branes, 
we see now a remarkable fact: scalar masses
larger
than gaugino masses can easily be obtained. This happens 
not only for narrow
ranges
of $\theta$. 
For example, for $e_O=1/3$ and $\theta_1=\pi/3$,
$\theta\approx 3\pi/2$ one obtains $m/|M|\approx 10$.
This result implies a relation of the type
$m_{\tilde l}\approx m_{\tilde q}\approx 3.5 M_{\tilde g}$.
An exhaustive analysis of other examples can be found in
\cite{Mio2}.
\bigskip

\noindent {\bf Acknowledgments}

\noindent  This work has been supported 
in part 
by the CICYT, under contract AEN97-1678-E, and
the European Union, under TMR contract ERBFMRX-CT96-0090.

\end{document}